\documentstyle [12pt]{article}

\newcommand{\be}{\begin{equation}}
\newcommand{\ee}{\end{equation}}
\newcommand{\bea}{\begin{eqnarray}}
\newcommand{\eea}{\end{eqnarray}}
\newcommand{\de}{\partial}
\newcommand{\nn}{\nonumber}

\newcommand{\da}{\dagger}

\newcommand{\refpa}[1]{(\ref{#1})}
\newcommand{\intll}[3]{\int _#1^#2 d\! #3 \,} 
\newcommand{\gh}{g \hbar}
\font\zz=cmss10
\newcommand{\Z}{\hbox{\zz Z} \kern-.4em \hbox{\zz Z}}
\begin{document}
\begin{titlepage}
\begin{flushleft}
G\"oteborg\\
ITP 93-33\\
hep-th/9308015\\
July 1993\\
\end{flushleft}
\vspace{1cm}
\begin{center}
{\Large $SU(N)$ Yang-Mills on a circle, loop variables, Mandelstam identities
and quantization ambiguities}\\
\vspace{5mm}
{\large Joakim Hallin}\footnote{Email address: tfejh@fy.chalmers.se}\\
\vspace{1cm}
{\sl Institute of Theoretical Physics\\
Chalmers University of Technology\\
and University of G\"oteborg\\
S-412 96 G\"oteborg, Sweden}\\
\vspace{1.5cm}
{\bf Abstract}\\
\end{center}
We consider $SU(N)$ Yang-Mills on a circle (cylindrical space-time) and
quantize the eigenvalues of the holonomy. In this
way
the Mandelstam identities associated with the holonomy are trivially solved.
Furthermore we indicate that there are exactly
two physically inequivalent representations of the algebra of gauge-invariant
operators, resulting in different spectra.
\end{titlepage}
\section{Introduction}
Wilson loop variables have been used for a long time in the study of Yang-Mills
theories. In recent years it has also
been used in the context of general relativity \cite{rs:qgr}. When
using the Wilson loop variable, which is the trace of the holonomy around a
closed loop, there always appear a number
of non-linear constraints, the Mandelstam identities, that encode that the
Wilson loop variable is really the trace of
an
$N\times N$-matrix. These identities are discussed in \cite{m:cm,rg,gt}. While
Yang-Mills theory in $3+1$-dimensions is
a difficult subject, Yang-Mills theory in $1+1$-dimensions is much simpler. In
fact it is almost completely trivial, having
only a finite number of global degrees of freedom on a circle. However, certain
aspects of the higher dimensional theories
remain, in particular the essence of the Mandelstam identities. Thus Yang-Mills
theory on a circle is a useful toy model.
It can be solved exactly. In \cite{hh:ym} and \cite{sgr:ym} this is
accomplished using other methods than ours. However, their
results differ which is discussed recently in \cite{h:cq}. We will also discuss
this difference. In fact, as we will indicate,
these are the only inequivalent ways of quantizing. In our formalism these two
inequivalent ways arise as different
representations of the algebra of gauge-invariant operators.

\section{Classical theory}
Our starting point is the gauge-invariant part of the $SU(N)$ Yang-Mills
Hamiltonian,
\be
H=\frac{1}{2}\intll{0}{L}{x} \mbox{tr}(E^2(x)),
\ee
where $L$ is the length of the circle.
The basic Poisson bracket is,
\be
\{ A^a(x),E^b(y) \}=\delta ^{ab} \delta (x-y),
\ee
and $A(x)=A^a(x) t^a$, $E(x)=E^a(x) t^a$ where $t^a$ are the $N\times N$-matrix
generators of the group ($a=1,\ldots ,
\mbox{dim}(SU(N))$). Here we have chosen
\[ \mbox{tr}(t^a t^b)=\delta ^{ab} \]
which implies the identity,
\be
(t^a)_{ij} (t^a)_{kl}=\delta_{il} \delta_{jk}-\frac{1}{N} \delta _{ij} \delta
_{kl},\label{eq:fierz}
\ee
where $i,j,k,l$ denotes matrix indices. We assume that the connection and
electric field are periodic fields on the
circle i.e. $A(x)=A(x+L)$, $E(x)=E(x+L)$.
We also have the first class constraint (Gauss' law),
\be
D_x E(x)=\de _x E(x)+ig \lbrack A(x),E(x) \rbrack \approx 0 ,\label{eq:gauss}
\ee
where $g$ is the coupling constant. Let us define parallel transport by
\be
U(x,y)={\cal{P}}\exp (ig \intll{x}{y}{x'}A(x')),
\ee
where $\cal{P}$ denotes path ordering, i.e. $U(x,y)$ is the solution to the
integral equation
\[
U(x,y)=1+ig \intll{x}{y}{x'}U(x,x') A(x').
\]
$U(x,y)$ is an element of the group $SU(N)$.
The holonomy $h(x)$ is defined by
\[ h(x)=U(x,x+L) \]
i.e. parallel transport once around the whole circle. Note also that $U(x,x+n
L)=h^n(x)$ where $n$ is an integer which
follows from the basic sewing property of path ordered exponentials,
\[ U(x,x') U(x',y)=U(x,y).\]
Let $\Lambda (x)$ be a (finite) $SU(N)$ gauge-transformation (generated by
\refpa{eq:gauss}). Then
\bea
A'(x) &=& \Lambda (x) A(x) \Lambda ^{-1}(x)+\frac{1}{ig}\Lambda (x) \de _x
\Lambda ^{-1}(x)\\
E'(x) &=& \Lambda (x) E(x) \Lambda^{-1}(x).
\eea
This implies that $U(x,y)$ transforms homogeneously i.e.
\be
U'(x,y)=\Lambda (x) U(x,y) \Lambda ^{-1}(y),
\ee
and in particular,
\be
h'(x)=\Lambda (x) h(x) \Lambda ^{-1}(x),\label{eq:hol}
\ee
if $\Lambda (x)$ is periodic. However,
this is not the most general situation. We might allow for non-periodic
gauge-transformations still keeping $A$ periodic.
These satisfy,
\be
\Lambda (x+L)=\Z _N \, \Lambda (x),
\ee
where $\Z _N$ is any element in the center of the group i.e. an $N$:th root of
unity $\xi$, $\xi ^N=1$. We will call such
non-periodic gauge-transformations $\Z _N$ transformations. Under such a
transformation, the holonomy transforms as,
\be
h'(x)=\xi \, \Lambda (x) h(x) \Lambda ^{-1}(x).
\ee
The $\Z _N$ symmetry is really of secondary importance since it only holds for
pure Yang-Mills. As soon as we couple fermions
this symmetry is lost.
\subsection{Loop variables}
Following \cite{rs:qgr} we introduce the following functions on phase space,
the loop variables,
\bea
T^0(n) &=& \mbox{tr}(h^n(x))\\
T^1(x;n) &=& \mbox{tr}(E(x) h^n(x))\\
T^2(x;n,y;m) &=& \mbox{tr}(E(x)U(x,y+nL)E(y)U(y,x+mL)).
\eea
They are easily seen to be gauge-invariant. Furthermore, $T^0(nN)$ etc, is $\Z
_N$ invariant.
Note also that $T^0(n)$ is independent of $x$, motivating the notation. In
fact, on the constraint surface $T^1(x;n)$ is also independent of $x$ since
\be
\de _x T^1(x;n)=\mbox{tr}((D_x E(x)) h^n(x))\approx 0 .\label{eq:t1}
\ee
Similarly, $T^2(x;n,y;m)$ is independent of $x$ and $y$. Using the identity,
\[ T^2(x+n' L;n,y;m)=T^2(x;n-n',y;m+n'), \]
it also follows that $T^2(x;n,y;m)$ is independent of $n-m$ on the constraint
surface, i.e. $T^2(x;n,y;m)=T^2(n+m)$.
Analogously, one may consider loop variables of higher order
in $E$. To calculate Poisson brackets we need,
\be
\frac{\delta U(x,y)}{\delta A^a(x')}=ig\theta (x,y,x')U(x,x')t^aU(x',y),
\ee
where
\[ \theta (x,y,x')=\intll{x}{y}{x''}\delta(x''-x').\]
In particular,
\be
\frac{\delta h(x)}{\delta A^a(x')}=igU(x,x')t^aU(x',x)h(x).\label{eq:func}
\ee
In what follows, all brackets will be evaluated on the constraint surface,
where they simplify. Using \refpa{eq:func}
and \refpa{eq:fierz} we obtain,
\bea
\{ T^0(n),T^0(m) \} &=& 0\nn \\
\{ T^1(n),T^1(m) \} &=& ig(n-m) T^1(n+m)-\frac{ig}{N}(n T^1(n) T^0(m)-m
T^1(m)T^0(n))\nn \\
\{ T^2(n),T^0(m) \} &=& -2igm (T^1(n+m)-\frac{1}{N}T^1(n) T^0(m))\nn \\
\{ T^2(n),T^1(m) \} &=& ig(n-2m)
T^2(n+m)+\frac{ig}{N}(2mT^1(n)T^1(m)-nT^2(n)T^0(m))\nn \\
\{ H,T^0(n) \} &=& -ignL\, T^1(n)\nn \\
\{ H,T^1(n) \} &=& -ignL\, T^2(n).\label{eq:pb2}
\eea
The last two identities follows since $H=\frac{L}{2}T^2(0)$. Brackets not
considered here lead to higher order
loop variables. We also have the reality conditions,
\bea
(T^0(n))^* &=& T^0(-n) \nn \\
(T^1(n))^* &=& T^1 (-n) \nn \\
(T^2(n))^* &=& T^2(-n) \nn \\
H^*=H,\label{eq:reality}
\eea
where $*$ denotes complex conjugation.

\subsection{Conjugacy classes}
As seen by \refpa{eq:hol}, the holonomy transforms under gauge-transformations
by conjugation in $SU(N)$. Gauge-invariant
functions of the holonomy are therefore class functions $f$,
\[ f(h)=f(g hg^{-1}), \, \, \, \, \, \forall g\in SU(N). \]
A particular example of a class function is $T^0(n)$. Let us note some
properties of the conjugacy classes of $SU(N)$, the
classic source of information being \cite{hw:cg}. Any $SU(N)$ matrix is
conjugate to a diagonal matrix $D$. Two diagonal matrices
are conjugate if and only if their eigenvalues are related by permutation. Let
$D=\mbox{diag}(\lambda _1,\ldots ,\lambda _N)$.
Since $\det D=1$ we have,
\be
\lambda _N=\lambda _1^{-1} \cdots \lambda _{N-1}^{-1}.\label{eq:ln}
\ee
Furthermore, since $D$ is unitary the eigenvalues all have modulus $1$ i.e.
$\lambda _i=e ^{i \varphi _i}$, ($\varphi _i$ real
$i=1,\ldots ,N-1$). Any class function $f$ is therefore a function of $N-1$
eigenvalues, symmetric under permutations
\[ \lambda _i\leftrightarrow \lambda _j, \, \, \, \, \, i,j=1,\ldots ,N\]
where $\lambda _N$ is given by \refpa{eq:ln}, e.g. for $N=2$, $f(\lambda
_1)=f(\lambda _1^{-1})$. From now on, permutations
will always mean permutations of all $N$ eigenvalues, $\lambda _N$ being given
by \refpa{eq:ln}. We can express $T^0(n)$ in
terms of the eigenvalues of $h(x)$ (which are independent of $x$),
\be
T^0(n)=\lambda _1^n+\ldots +\lambda _{N-1}^n+\lambda _1^{-n} \cdots \lambda
_{N-1}^{-n}.
\ee

\section{Quantization}
In the Dirac quantization approach, the quantized constraint operators should
annihilate physical states. Therefore, by
\refpa{eq:t1},
\[ \de _x\hat T^1(x;n) \Psi _{phys}=0,\]
i.e. $\hat T^1(x;n)=\hat T^1(n)$ and similarly for $\hat T^2$ (or rather, $\Psi
_{phys}$ has support only on $\hat T^1(x;n)$:s that
are independent of $x$). Hence quantize the Poisson bracket algebra
\refpa{eq:pb2} as,
\bea
\lbrack \hat T^0(n),\hat T^0(m) \rbrack &=& 0\label{eq:t0t0}\\
\lbrack \hat T^1(n),\hat T^0(m) \rbrack &=& \gh m (\hat
T^0(n+m)-\frac{1}{N}\hat T^0(n) \hat T^0(m))\label{eq:t1t0}\\
\lbrack \hat T^1(n),\hat T^1(m) \rbrack &=& \gh (m-n)\hat T^1(n+m)+\frac{\gh
}{N}(n \hat T^0(m) \hat T^1(n)-
	m \hat T^0(n) \hat T^1(m))\label{eq:t1t1}\\
\lbrack \hat H,\hat T^0(n) \rbrack &=& \gh L n\, \hat T^1(n)\label{eq:ht0}\\
\lbrack \hat H,\hat T^1(n) \rbrack &=& \gh L n\, \hat T^2(n).\label{eq:ht1}
\eea
We have refrained from quantizing the algebra involving $\hat T^2$ since it is
ordering dependent. Instead we define
$\hat T^2(n)$ by \refpa{eq:ht1}. Assuming $T^2(n)$ to be continuous in $n$ we
obtain,
\be
\lim _{n\rightarrow 0}\frac{1}{2\gh n}\lbrack \hat H,\hat T^1(n) \rbrack =\hat
H.\label{eq:hlim}
\ee
Note that the last two terms in \refpa{eq:t1t1} look ordering dependent, but in
fact they are not due to \refpa{eq:t1t0} as long
as one orders both $\hat T^1$:s either to the left or to the right of the $\hat
T^0$:s. Now let these
operators act on
wavefunctions that are class functions, i.e. symmetric functions of $N-1$
eigenvalues of the holonomy $h(x)$. The eigenvalue
operators themselves (which are not gauge-invariant), act simply by
multiplication and hence $\hat T^0(n)$ acts as,
\be
\hat T^0(n) \Psi (\lambda _1,\ldots ,\lambda _{N-1})=(\lambda _1^n+\ldots
+\lambda _{N-1}^n+\lambda _1^{-n} \cdots
\lambda _{N-1}^{-n}) \Psi (\lambda _1,\ldots ,\lambda _{N-1}).
\ee
This ensures that \refpa{eq:t0t0} is satisfied. Furthermore, this
representation of $\hat T^0$ automatically satisfies the
Mandelstam identities, e.g. when $N=2$,
\bea
\hat T^0(n) &=& \hat T^0(-n)\nn \\
\hat T^0(n) \hat T^0(m) &=& \hat T^0(n+m)+\hat T^0(n-m).\nn
\eea
Under a $\Z _N$ transformation, $\Psi (\lambda _1,\ldots ,\lambda _{N-1})$
transforms into $\Psi (\xi \lambda _1,\ldots ,\xi
\lambda _{N-1})$, where $\xi ^N=1$.  Let's try to find a representation of
$\hat T^1$ satisfying \refpa{eq:t1t0} as a
(pure) first order differential operator in the eigenvalues. The unique
solution is, labeling this specific representation
by $\hat T^1_0$,
\be
\hat T^1_0(n) \Psi (\{ \lambda \} )=\gh (\sum _{i=1}^{N-1}(\lambda
_i^n-\frac{1}{N}\hat T^0(n)) \lambda _i \de _{\lambda _i})
\Psi (\{ \lambda\} ),\label{eq:mint1}
\ee
where $\{ \lambda \} =(\lambda _1,\ldots ,\lambda _{N-1})$. A check shows that
this representation of $\hat T^1$ is invariant
under permutations (of the eigenvalues) and that it satisfies \refpa{eq:t1t1}.
However, we might add a zeroth order term to
this representation, i.e. a class function, as long as \refpa{eq:t1t1} is
satisfied (\refpa{eq:t1t0} is obviously still satisfied).
We will investigate such choices in what follows. First however consider
\refpa{eq:ht0}. As an ansatz assume that $\hat H$ is
a (pure) second order differential operator in angles $\varphi _i$, i.e. a
(pure) second order polynomial in $\lambda _i
\de _{\lambda _i}$. Then compare only the first order part of the left- and
righthand sides of \refpa{eq:ht0}. The unique solution
is given by,
\be
\hat H_0 \Psi (\{ \lambda\} )=\frac{(\gh )^2 L}{2}(\sum _{i=1}^{N-1} (\lambda
_i \de _{\lambda _i})^2-\frac{1}{N}
(\sum _{i=1}^{N-1} \lambda _i \de _{\lambda _i})^2)\Psi (\{ \lambda\} ).
\ee
Let us note some properties of $\hat H_0$. Introduce
\be
\Xi _{(n_1,\cdots ,n_{N-1})}(\{ \lambda \} )=\lambda _1^{n_1} \cdots \lambda
_{N-1}^{n_{N-1}},
\ee
where $n_1,\ldots ,n_{N-1}$ are integers. $\Xi$ is an eigenvector of $\hat
H_0$, i.e.
\be
\hat H_0 \, \Xi _{(n_1,\cdots ,n_{N-1})}(\{ \lambda \} )=\frac{ (\gh )^2 L}{2N}
P_N(n_1,\ldots ,n_{N-1})
\Xi _{(n_1,\cdots ,n_{N-1})}(\{ \lambda \} ),\label{eq:eigen}
\ee
where
\be
P_N(\{ n\} )=(N-1) \sum _{i=1}^{N-1} n_i^2-2\sum _{j>i=1}^{N-1} n_i n_j .
\ee
To $\hat H_0$ we might add a first order term. We will return to this point in
a while.

\subsection{Symmetric representation}
Let us choose $\hat H=\hat H_0$. Checking, it is found that it satisfies
\refpa{eq:hlim}. This is in fact, in our formalism,
 the Hamiltonian derived in \cite{hh:ym}. $\hat T^1$ and $\hat T^2$ are
determined by \refpa{eq:ht0} and \refpa{eq:ht1} respectively. The
eigenstates are totally symmetric
linear combinations of $\Xi _{ ( \{ n \}) }$ (remember that physical states are
class functions), i.e.
\[ \Psi _{(n_1,\ldots ,n_{N-1})} (\{ \lambda \} )=\sum _{perms} \Xi
_{(n_1,\ldots ,n_{N-1})}(\pi (\lambda _1),\ldots ,\pi (\lambda _{N-1})), \]
where $\pi$ permutes all $\lambda _i$:s including $\lambda _N$. Evidently, not
all indices $(n_1,\ldots ,n_{N-1})$ correspond to different
eigenstates. If we want these states to be $\Z _N$ invariant we have to require
$\sum _{i=1}^{N-1}n_i$ to be a multiple of
$N$. The eigenenergies are given by \refpa{eq:eigen}. The action of the loop
variables is very simple on the eigenstates, e.g.
\[
\hat T^0(n)\Psi _{(n_1,\ldots ,n_{N-1})}(\{ \lambda \})=\sum _{i=1}^{N-1}\Psi
_{(n_1,\ldots ,n_i+n,\ldots ,n_{N-1})}(\{ \lambda \})+
\Psi _{(n_1-n,\ldots ,n_{N-1}-n)} (\{ \lambda \}).
\]
An inner product is determined by requiring $(\hat T^0 (n))^\da =\hat T^0(-n)$
and $\hat H^\da =\hat H$. Then \refpa{eq:ht0} and
\refpa{eq:ht1} implies that all the classical reality conditions,
\refpa{eq:reality}, are quantized exactly.
Hence, (up to an overall factor),
\be
< \Phi ,\Psi >=\int d\varphi _1 \cdots d\varphi _{N-1} \Phi ^* (\{ \varphi \}
)\Psi (\{ \varphi \} ).
\ee
Here all integrals are taken from $-\pi$ to $\pi$ in the angles. Alternatively
we can integrate over the eigenvalues,
\[ \frac{d\lambda _i}{i \lambda _i}=d\varphi _i .\]
Different eigenstates are orthogonal using this inner product. The groundstate
is $\Psi _{(0,\ldots ,0)}$ and it has zero energy.

\subsection{Antisymmetric representation}
Now choose $\hat H=\Delta ^{-1} \hat H_0 \Delta$ where $\Delta$ is,
\[ \Delta =\prod _{j>i=1}^N(\lambda _i-\lambda _j).\]
We note that it is a well defined choice, being invariant under permutations of
eigenvalues. Pulling $\hat H_0$ through to
the right, one sees that it corresponds to having added a certain first order
term to $\hat H_0$. Furthermore, it satisfies
\refpa{eq:hlim}. This is the Hamiltonian considered in \cite{sgr:ym}. It is (up
to a constant) the radial part of the
Laplacian on $SU(N)$, see \cite{hel}. $\Delta$ is totally
antisymmetric under permutations of eigenvalues. Hence eigenstates of
$\hat H$ are given as,
\[ \Psi _{(n_1,\ldots ,n_{N-1})} (\{ \lambda \} )=\Delta ^{-1} \sum _{perms}
\mbox{sgn}(\pi)\,  \Xi _{(n_1,\ldots ,n_{N-1})}(\pi (\lambda _1),\ldots ,
\pi (\lambda _{N-1})). \]
These are the characters of $SU(N)$. Eigenenergies are still given by
\refpa{eq:eigen}. The groundstate is
$\Psi _{(1,\ldots ,N-1)}$ with energy
\[ (\gh )^2L\frac{N}{24}(N^2-1).\]
The
spectrum of $\hat H$ is a proper subset of that of $\hat H_0$. Hence these
Hamiltonians are clearly physically inequivalent. The action of
loop variables on eigenstates is the same as for the symmetric representation.
The inner product
is,
\[ < \Phi ,\Psi >=\int d\varphi _1 \cdots d\varphi _{N-1} \Delta \Delta ^* \Phi
^* (\{ \varphi \} )\Psi (\{ \varphi \} ).\]
The measure density $\Delta \Delta ^*$ is the measure density induced by the
Haar-measure on the group. Note how utterly sensible it is
from the point of view
of the group, e.g. the conjugacy class
$\lambda _1=\ldots =\lambda _{N-1}=1$ consists of a single group element, the
unit matrix, in contrast to a generic conjugacy class having all
eigenvalues distinct which consists of a set of group elements forming a
submanifold of the group with non-zero dimension. Thinking about
the group it is natural to give a larger weight to this generic conjugacy class
than the unit element class. $\Delta$ does just this as it vanishes
on the unit element class. In general, the so called singular set which is the
set of conjugacy classes having not all eigenvalues distinct, has
Haar-measure zero ($\Delta$ is zero on this set).

\subsection{Generalities}
Returning to the issue of adding first order terms to $\hat H_0$ (of which the
antisymmetric representation is a particular example), consider
for simplicity $N=2$. In this case the most general representation
also satisfying \refpa{eq:hlim} is,
\be
\hat H \Psi (\varphi _1 )=-\frac{ (\gh )^2 L}{4}(\de _{\varphi _1}^2+2f(\varphi
_1)\de _{\varphi _1}+f(\varphi _1)^2+f'(\varphi _1))
\Psi (\varphi _1),\label{eq:mostgen}
\ee
where $f(-\varphi _1)=-f(\varphi _1)$ to make $\hat H$ gauge-invariant, i.e.
invariant under permutations of eigenvalues. Defining $\hat T^2$ by
\refpa{eq:ht1} implies that $\hat T^2$ satisfies its quantized
bracket algebra with a
particular ordering, modulo some quantum corrections in $\lbrack \hat T^2,\hat
T^1 \rbrack$. These corrections are
independent of $f(\varphi _1)$. This is not surprising as we will see. In the
language of \cite{aa:ct} the Hamiltonian in
\refpa{eq:mostgen} and $\hat H_0$ are related by a quantum canonical
transformation. The measure density in the inner product determined by
this Hamiltonian is found to be,
\[ \mu (\varphi _1)=k e^{2F(\varphi _1)}, \]
where $k$ is a constant and $F'(\varphi _1 )=f(\varphi _1)$. From the measure
density we can construct a quantum canonical transformation by
letting,
\be
\mu (\varphi _1)=C^{-2}(\varphi _1).\label{eq:can}
\ee
A solution is $C(\varphi _1)=e^{-F(\varphi _1)}$ having set $k=1$. It is easily
seen that $\hat H=C\, \hat H_0 \, C^{-1}$
and hence $\hat H$ and $\hat H_0$ are related by a canonical transformation. It
would seem that therefore $\hat H$ and
$\hat H_0$ are physically equivalent. There is however a subtlety. States are
required to be invariant under permutations of
eigenvalues, which in this case means even in $\varphi _1$. We have chosen a
$C(\varphi _1)$ which is even, hence eigenstates
of $\hat H$ are,
\[ \Psi (\varphi _1)=C(\varphi _1) \cos n  \varphi _1.\]
Thus in this case, $\hat H$ and $\hat H_0$ are completely equivalent. We might
however have chosen a different
square root in \refpa{eq:can} such that $C(\varphi _1)$ is odd. For such a $C$,
eigenstates of $\hat H$ are,
\[ \Psi (\varphi _1)=C(\varphi _1) \sin n  \varphi _1.\]
This is exactly what happens in the antisymmetric representation, having
$f(\varphi _1)=\cot \varphi _1$. The measure
is,
\[ \mu (\varphi _1)=e^{2\log |\sin \varphi _1|}=\sin ^2\varphi _1.\]
A possible choice of $C$ is obviously $1/ \sin \varphi _1$ which is an odd
function of $\varphi _1$. Hence
eigenstates of the Hamiltonian $\sin ^{-1}\varphi _1\hat H_0 \sin \varphi _1$
has eigenstates $\frac{\sin n
\varphi _1}{\sin \varphi _1}$. It is thus clear that all choices of
Hamiltonians are either equivalent to the symmetric
or the antisymmetric representation and that this quantization ambiguity arises
when choosing a particular square root in \refpa{eq:can}. This conclusion
generalizes to higher $N$. For any $N$, we can
write the most general Hamiltonian $\hat H$ as,
\[ \hat H=\hat H_0+\sum _{i=1}^{N-1}f_i(\{ \lambda \})\lambda _i \de _{\lambda
_i}+g(\{ \lambda \}),\]
where $g(\{ \lambda \})$ is determined from \refpa{eq:hlim}. The functions
$f_i$ are not independent as the Hamiltonian
is required to be invariant under permutations. This Hamiltonian determines a
certain measure density $\mu (\{ \lambda
\})$ which is invariant. Setting $\mu =C^{-2}$ determines $C$ up to a sign (a
sign which can vary from point to point).
We can e.g. choose $C$ either totally symmetric or totally antisymmetric. Then
everything works fine. There are however
other choices of $C$ which are neither totally symmetric nor totally
antisymmetric. But in these cases it seems impossible
to construct eigenfunctions of the Hamiltonian which are totally symmetric.
Thus we must exclude these choices i.e. in
general $\hat H=C\, \hat H_0\, C^{-1}$ and this Hamiltonian is
equivalent to either the symmetric or the antisymmetric representation.

\section{Conclusion}
We have seen that considering the eigenvalues of the holonomy is very fruitful
when quantizing Yang-Mills theory. This
feature is expected to generalize to higher dimensions. The quantization
ambiguity is seen to arise as follows. Even if
two different Hamiltonians are related by a quantum canonical transformation,
their eigenstates might not be related by
this canonical transformation since eigenstates are required to be invariant
under permutations of eigenvalues. As far as
pure Yang-Mills theory goes, there is no good reason to reason to prefer one
representation to the other. However we can
speculate that coupling (fundamental) fermions might change matters since then
the whole group becomes important and not
just the conjugacy classes.\\
\\
We wish to thank B. Nilsson and B.S. Skagerstam for discussions.

\end{document}